\documentclass[useAMS,usenatbib]{mn2e}
\usepackage{url}
\usepackage{graphicx}
\usepackage{verbatim}
\usepackage[usenames]{color}
\usepackage[flushleft]{threeparttable}
\DeclareGraphicsExtensions{.pdf,.png,.jpg,.mps,.eps,.ps}
\usepackage{amsmath}
\usepackage{natbib}
\usepackage{color}
\usepackage{graphicx}
\usepackage{lscape}
\usepackage{gensymb}
\usepackage{amssymb}
\usepackage{psfrag}

\newcommand\nicer{{\it NICER}}
\newcommand\nustar{{\it NuSTAR}}
\newcommand\swift{{\it SWIFT}}

\newcommand\chandra{{\it Chandra}}

\newcommand\rxte{{\it RXTE}}

\newcommand\inte{{\it INTEGRAL}}

\newcommand\kev{{\rm~keV}}

\newcommand\kms{\ifmmode {\rm~km\ s}^{-1} \else ~km s$^{-1}$\fi}
\newcommand\Hunit{\ifmmode {\rm~km\ s}^{-1}\ {\rm Mpc}^{-1}
        \else ~km s$^{-1}$ Mpc$^{-1}$\fi}
\newcommand\ctssec{\ifmmode {\rm~count\ s}^{-1} \else ~count s$^{-1}$\fi}
\newcommand\ergsec{\ifmmode {\rm~erg\ s}^{-1} \else
        ~erg s$^{-1}$\fi}
\newcommand\funit{\ifmmode {\rm~erg\ s}^{-1}\;{\rm cm}^{-2} \else
        ~ergs s$^{-1}$ cm$^{-2}$\fi}
\newcommand\phflux{\ifmmode {\rm~photon\ s}^{-1}\;{\rm cm}^{-2}
        \else   ~photon s$^{-1}$ cm$^{-2}$\fi}
\newcommand\efluxA{\ifmmode {\rm~erg\ s}^{-1}\;{\rm cm}^{-2}\;{\rm
        \AA}^{-1} \else ~erg s$^{-1}$ cm$^{-2}$ \AA$^{-1}$\fi}
\newcommand\efluxHz{\ifmmode {\rm~erg\ s}^{-1}\;{\rm cm}^{-2}\;{\rm
        Hz}^{-1} \else ~erg s$^{-1}$ cm$^{-2}$ Hz$^{-1}$\fi}
\newcommand\cc{\ifmmode {\rm~cm}^{-3} \else cm$^{-3}$\fi}
\newcommand\FWHM{\ifmmode {\rm~FWHM} \else ${\rm~FWHM}$\fi}
\newcommand\Msun{\ifmmode M_{\odot} \else $M_{\odot}$\fi}
\newcommand\Lsun{\ifmmode L_{\odot} \else $L_{\odot}$\fi}

\newcommand\hbeta{\ifmmode {\rm H}\beta \else H$\beta$\fi}
\newcommand\Kalpha{\ifmmode {\rm K}\alpha \else K$\alpha$\fi}
\newcommand\nh{\ifmmode N_{\rm H} \else N$_{\rm H}$\fi}

\usepackage{graphicx}
\usepackage{url}
\usepackage{verbatim}
\usepackage{color}

\title[Disc reflection and absorption in 1A~1744-361]{Relativistic X-ray reflection and highly ionized absorption in the spectrum of NS LMXB 1A~1744-361}
\author[Mondal et al.]{\parbox[]{6.5in}{Aditya S. Mondal$^{1}\thanks{E-mail: adityas.mondal@visva-bharati.ac.in}$, B. Raychaudhuri$^{1}$, G. C. Dewangan$^{2}$   \\
\small
$^{1}$Department of physics, Visva-Bharati, Santiniketan, West Bengal-731235, India \\
$^{2}$Inter-University Centre for  Astronomy \& Astrophysics (IUCAA), Pune, 411007 India \\
}}

\date{\today}
\begin{document}
\maketitle
\begin{abstract}
We present the results from the spectral and timing analysis of the accreting neutron star 1A~1744-361 from the \nustar{} observation performed in its 2022 outbursts. The unabsorbed bolometric X-ray luminosity during this observation in the energy band $0.1-100\kev{}$ is $3.9\times 10^{37}$ erg~s$^{-1}$, assuming a distance of $9$ kpc. During this observation, the source was in the banana branch of the atoll track. The source spectrum exhibits relativistic disc reflection and clear absorption features when an absorbed blackbody and cut-off power-law model describes the continuum emission. The $3-50\kev{}$ source spectrum is well fitted using a model combination consisting of an absorbed single-temperature blackbody and a reflection model along with the addition of a warm absorber component. The inner-disk radius, $R_{in}$, obtained from the reflection fit is $\sim(1.61-2.86)R_{ISCO}=(8.4-14.9)R_{g}$ ($17.6-31.2$ km for a $1.4\Msun$ NS). This measurement allowed us to further constrain the magnetic field strength to $B\lesssim 0.94\times 10^{9}$G. The strong absorption features $\sim 6.91\kev{}$ and $\sim 7.99\kev{}$ imply the presence of highly ionized absorbing material with a column density $N_{H}$ of $\sim 3\times 10^{22}$ cm$^{-2}$, emanating from the accretion disk in the form of disc wind with an outflow velocity of $v_{out}\simeq 0.021c\simeq 6300$ km s$^{-1}$.
 
\end{abstract}

\section{Introduction}
A Neutron star Low-mass X-ray binary (NS LMXB) consists of an NS and a low-mass ($\lesssim 1 \Msun$) donor star. In NS LMXB systems, mass accreting from the donor star onto the NS by Roche-lobe overflow forms structures such as an accretion disc. It emits electromagnetic radiation in a wide range of wavelengths, from radio to X-rays. NS LMXBs may be persistent accreator or transient systems based on their long-term variabilities. Persistent NS LMXBs are characterized by persistent luminosity in X-rays and may have an X-ray luminosity of $L_{X}\gtrsim 10^{36} \ergsec{}$ (\citealt{2019ApJ...873...99L, 2017ApJ...836..140L}). Whereas some NS LMXBs exhibit an outburst, a sudden and explosive brightening phenomenon, and such LMXBs are called X-ray Transients. Transient LMXBs undergo recurrent bright ($L_{X}\gtrsim 10^{36} \ergsec{}$) outbursts lasting from days to weeks and then return to long intervals of X-ray quiescence ($L_{X}\lesssim 10^{34} \ergsec{}$) lasting from months to years \citep{2010A&A...524A..69D}. Persistent and transient NS LMXBs are classified into two classes: {\it Z} and {\it Atoll} sources based on their behavior on the X-ray hardness intensity diagram (HID) and color-color diagram (CCD) \citep{1989A&A...225...79H}. {\it Z} sources are usually very bright and sometimes radiate at Eddington luminosity ($L_{Edd}$). {\it Atoll} sources are generally less bright ($L\lesssim 0.5 L_{Edd}$). Based on their luminosity, the state of {\it Atoll} sources is further divided into ‘banana branch’ and ‘island branch’, and usually, they correspond to the High-soft (HS) and Low-hard (LH) state, respectively. The banana branch has been further divided into lower banana and upper banana at lower and higher luminosities. \\

The source 1A~1744-361 is a transient NS LMXB discovered by the \textit{Ariel V} satellite in 1976 during its outburst state \citep{1976IAUC.2925....2D, 1977MNRAS.179P..27C}. Since then, several outbursts have been observed from the source between the years 1989 and 2005 with a number of missions like \rxte{}, \chandra{}, and \inte{}. \citet{2001AstL...27..781E} discovered 33 likely type-I bursts in the field containing 1A~1744-361 in 2001. However, those were not unambiguously identified as thermonuclear bursts. At a later time, the first thermonuclear (Type I) burst from 1A~1744-361 was discovered by \citet{2006ApJ...639L..31B} using the 2005 \rxte{} PCA data. It confirmed the suggestion of \citet{2001AstL...27..781E} that the source harbors a rapidly rotating NS as a compact object. During this burst \citet{2006ApJ...639L..31B} discovered millisecond period brightness oscillation, which provided the spin frequency of the NS of $\sim 530$ Hz. The lack of strong indication of photospheric radius expansion during the burst suggested a 9 kpc upper limit of the source distance \citep{2006ApJ...639L..31B}. From the studies of energy-dependent dips in the 2003 PCA data \citet{2006ApJ...639L..31B} also found that this source is a dipping LMXB. The binary orbital period of the source estimated by \citet{2006ApJ...639L..31B} is $ 97\pm 22$ minutes. From the timing analysis, \citet{2006ApJ...652..603B} found that the source 1A~1744-361 shows {\it Atoll} behavior during the outbursts. After July 2005, the source's outburst activities were detected on June 2008, November 2009, and August 2013 \citep{2013ATel.5301....1B}. Recently, MAXI/GSC detected an outburst from the field, including 1A~1744-361 on 30th May 2022 \citep{2022ATel15407....1K}. \swift{}/XRT subsequently confirmed the outburst originating from the known neutron star low-mass X-ray binary 1A~1744-361 \citep{2022ATel15408....1K}. At later times, the source was observed by the missions \nicer{} and \nustar{} during 3rd June 2022 and 8th June 2022, respectively \citep{2022ATel15424....1N, 2022ATel15429....1P}.\\

Spectral studies of 1A~1744-361 were performed in the past using data from the satellites \rxte{} and \chandra{} \citep{2006ApJ...652..603B, 2012ApJ...753....2G}. \citet{2006ApJ...652..603B} performed the spectral analysis of this source using the \rxte{}/PCA data, and they found that a Comptonized blackbody model describes the persistent spectrum of this source well. They reported the first detection of a broad ($\sim 0.6 \kev{}$) iron emission feature at $\sim 6$ keV and an iron absorption edge at $\sim 8$ keV. They mentioned that the energies of the absorption edge are consistent with those expected from ionized iron. \citet{2012ApJ...753....2G} reported on the \chandra{} High Energy Transmission Grating (HETG) spectra of this source during its 2008 July outburst. They found that its persistent emission is well modeled by a blackbody ($kT\sim 1.0\kev{}$) plus power law ($\Gamma \sim 1.7$) with an absorption edge. They also found a significant absorption line at $6.961 \pm 0.002$ keV, consistent with the Fe XXVI (hydrogen-like Fe) $2-1$ transition. They placed an upper limit on the velocity of a redshifted flow of $v < 221$ km s$^{-1}$. They further confirmed the suggestion of \citet{2006ApJ...652..603B} that this source is an {\it Atoll} source. The mission \nustar{} observed the source on 8th June 2022, and the preliminary spectral results have been reported by \citet{2022ATel15429....1P}. They found that the $3.0-40.0\kev{}$ energy spectrum is well-described by a combination of relativistic disk reflection of a power law with a high energy cut-off, with an intervening warm absorber. They also reported the detection of narrow absorption lines at $6.95 \pm 0.02$ keV and $8.01 \pm 0.03$ keV. The time-averaged $0.5-10.0$ keV \nicer{} spectrum during its 2022 outburst is well-fitted by an absorbed cut-off power law and blackbody component with a Gaussian absorption line component \citep{2022ATel15424....1N}. The Gaussian absorption line has centroid energy $6.988^{+0.012}_{-0.015}$ keV and width $\sim 0.004 \kev{}$ .\\

In this work, we represent a systematic and in-depth spectral and timing analysis of this source using the available \nustar{} observation performed on 8th June 2022. We aim to characterize the shape of the X-ray spectrum accurately and search for any emission and/or absorption features in it, as there was a hint of the same from the previous analysis. Studies of the same allow us to derive information on the physical and geometrical parameters of the system. We have organized the paper as follows: in Section 2 we describe the observation and data reduction. In Section 3 and Section 4 we present the details of the timing and spectral analysis, respectively. In section 5, we discuss the results obtained from the analysis.

\begin{figure*}
\centering
\includegraphics[scale=0.50, angle=0]{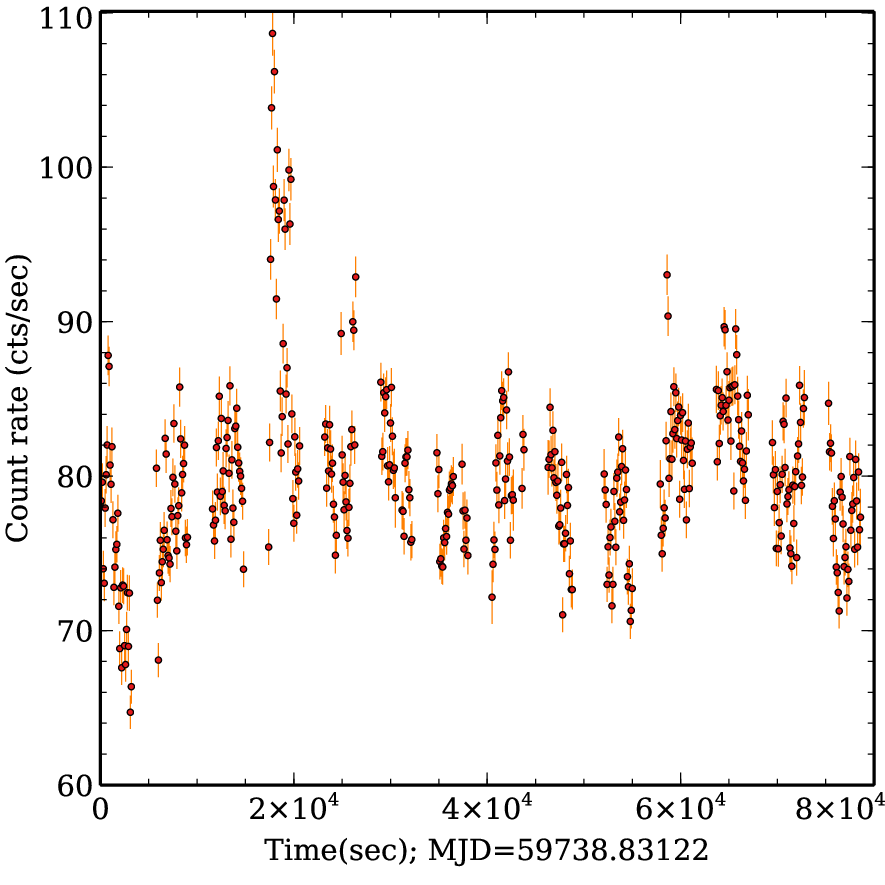}\hspace{2cm}
\includegraphics[scale=0.50, angle=0]{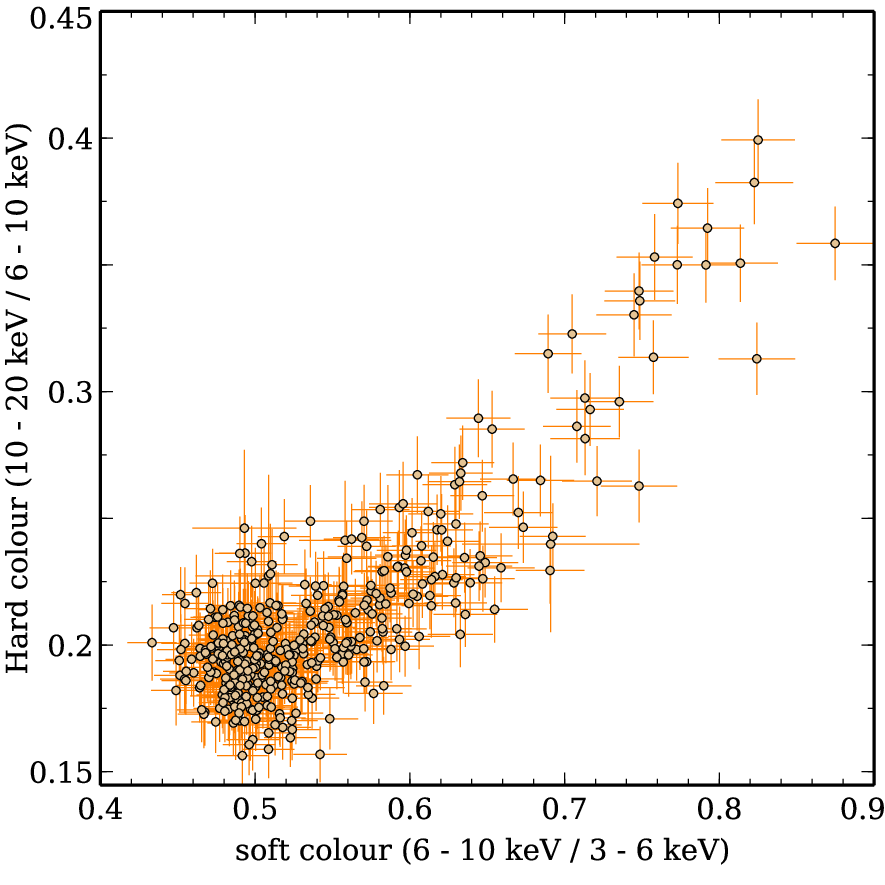}
\caption{Left panel: $3-79\kev{}$ \nustar{}/FPMA light curve of the source with a binning of 100s. The source shows a small variations in intensity during this observation. Right panel: the Color-color diagram of the source that is consistent with the {\it banana} branch of the {\it Atoll} sources.} 
\label{Fig1}
\end{figure*}

\section{observation and data reduction}
\textit{Nuclear Spectroscopic Telescope ARray} (\nustar{}; \citealt{2013ApJ...770..103H}) observed the source 1A~1744-361 only once on 8th June 2022 for a total exposure of $\sim 34$ ks. We have used this observation (obsID: 90801312001) for our analysis.  \\

The \nustar{} data were collected in the $3-79$\kev{} energy band using two identical co-aligned telescopes equipped with the focal plane modules FPMA and FPMB. We reduced the data using the standard data analysis software {\tt NUSTARDAS v2.1.1} task included in {\tt HEASOFT v6.29} and using the latest {\tt CALDB} version available. We used the task {\tt nupipeline} to generate the calibrated and screened event files. 
For the source, we extracted events from a circular region centered on the source, with a radius of $100''$ for both the FPMA and FPMB.
We also extracted background events from a circle of the same radius from the area of the same chip with the lowest apparent source contamination for both instruments. We then used the tool {\tt nuproducts} to build the filtered event files, the background subtracted light curves, the spectra, and the arf and rmf files. This {\tt nuproducts} software ensures that all the instrumental effects, including loss of exposure due to dead-time, are correctly accounted for. We grouped the FPMA and FPMB spectral data with a minimum of $50$ counts per bin which allows the use of $\chi^{2}$ statistics. Finally, we fitted spectra from the FPMA and FPMB simultaneously, leaving a floating cross-normalization constant.

\begin{figure}
\centering
\includegraphics[scale=0.32, angle=-90]{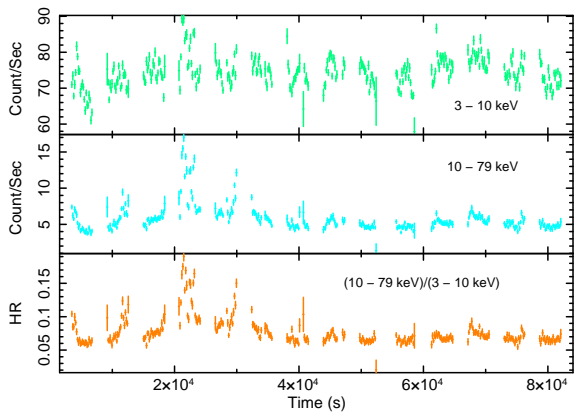}
\caption{The top and the middle panels show the source count rate with time in the $3-10$ \kev{} and $10-79$ \kev{} energy bands, respectively. The bottom panel shows the variation of the hardness ratio which is defined here as the $10-79$ \kev{} count rate divided by the $3-10$ \kev{} count rate with time.} 
\label{Fig2}
\end{figure}

\section{Temporal Analysis}
The left panel of Figure~\ref{Fig1} shows the 100s bin-size \nustar{} light curve of the source in the $3-79$ \kev{} energy band. The light curve exhibits variations in intensity throughout the observation and a short flare at an elapsed time around $20$ ks. The variability of the X-ray flux is commonly explained by the partial covering of the NS system by ionized absorbers \citep{2021MNRAS.503.5600B, 2023A&A...674A.100D, 2023MNRAS.522.3367S}. We did not observe either Type-I X-ray bursts or absorption dips in the \nustar{} light curve of this source, although those have been detected earlier \citep{2006ApJ...639L..31B}. The average count rate of the source during this observation is $\sim 80$\ctssec{}, while a small increment (factor of $\sim 1.4$) in the count rate is observed during the flaring. We further constructed the colour-colour diagram (CCD) of the source using light curves in the $3-6$\kev{}, $6-10$\kev{}, and $10-20$\kev{} energy bands. We defined the soft-colour as the ratio of count rates in $6-10$\kev{} and $3-6$\kev{}, and hard colour as the ratio of count rates in $10-20$\kev{} and $6-10$\kev{}. We present the resulting diagram in the right panel of Figure~\ref{Fig1}. It is evident from the CCD that the source is in the 'upper-banana' state during this observation. The CCD of this source is very similar to the other known {\it Atoll} source 4U~1608-522 and GX~9+9 \citep{2011ApJ...738...62T, 2023MNRAS.521L..74C}. In addition, we extracted light curves with 100s bins in the $3-10$ \kev{} and $10-79$ \kev{} energy bands and calculated the hardness ratio (HR). The variation of the $3-10$\kev{} and $10-79$\kev{} count rates and HR as a function of time are shown in the top, middle, and lower panel of Figure~\ref{Fig2}, respectively. We note that the HR remains roughly constant with time throughout this observation except for the short time interval near the elapsed time of $\sim 20$ ks where flaring occurs. It further establishes the commonly observed fact that in the banana state of the {\it Atoll} sources, the hardness is relatively constant over a wide range of luminosity \citep{1995xrbi.nasa..252V, 2014MNRAS.438.2784C}. \\

\begin{figure*}
\includegraphics[scale=0.48, angle=-90]{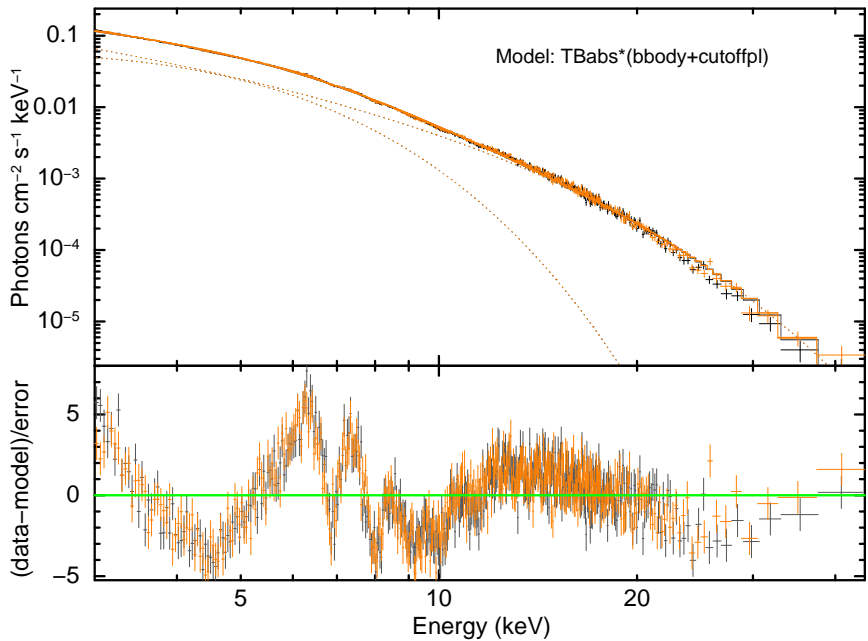}
\caption{The source spectrum in the energy band $3.0-50.0\kev{}$ obtained from the \nustar{} FPMA and FPMB is presented here. The continuum emission is fitted with the absorbed {\tt bbody} and the  {\tt cutoffpl} model, shown with the dotted lines. Fit ratio associated with this continuum model is shown in the bottom panel of this plot. The presence of disc reflection and clear absorption features are evident in the spectrum.} 
\label{Fig4}
\end{figure*}

\section{spectral analysis}
We used the spectral analysis package {\tt XSPEC} $v12.12.0$ \citep{1996ASPC..101...17A} to fit the \nustar{} FPMA and FPMB spectra simultaneously between $3.0$ to $50.0$ \kev{} energy band, while above $50$\kev{} is dominated by the background. We added a model {\tt constant} to account for cross-calibration of the two instruments FPMA and FPMB. The value of {\tt constant} for FPMA was fixed to $1$ and allowed it to vary for the FPMB. We used {\tt TBabs} to model the Galactic absorption along the line of sight with {\tt wilm} abundances \citep{2000ApJ...542..914W} and {\tt vern} \citep{1996ApJ...465..487V} photoelectric cross section. Spectral uncertainties are quoted at the $90$ percent confidence intervals (e.g., $\Delta\chi^{2}=2.7$ for one interesting parameter) unless otherwise stated.\\

\subsection{Continuum modeling}
In order to probe the spectral shape of the source, we started by fitting the \nustar{} spectrum with simple phenomenological models. We first tried a combination of an absorbed power-law ({\tt cutoffpl}) component and a single-temperature blackbody component ({\tt bbody}). The {\tt cutoffpl} model is used to describe the Comptonized emission (nonthermal) from the corona, and the {\tt bbody} component accounts for the thermal emission from the NS surface/boundary layer. This combination of models has previously been used by \citet{2012ApJ...753....2G} for this source and by many authors to model the continuum emission successfully for atoll sources (\citealt{2004A&A...426..979F, 2007ApJ...667.1073L}). We found that this two-component model approximates the continuum reasonably well, the poor quality of the fit ($\chi^2/dof=3712/849$) is mainly due to the presence of strong reflection features. We obtained a blackbody temperature $kT_{bb}\sim 1.21\kev{}$, a power-law photon index $\Gamma\sim 1.19$ and the cut-off energy $E_{cut}\sim 4.95$\kev{}. The observed cut-off energy is typical for the atoll sources in the banana states \citep{2014MNRAS.438.2784C}. We also noted that the Comptonized emission is the dominant emission comprising $\sim 60\%$ of the total $3-70\kev{}$ flux. The residual of this model is shown in Figure~\ref{Fig4}. The fit residual reveals the presence of several narrow emission and/or absorption features in $\sim 6-9$\kev{}, also an indication of an asymmetric Ke K line profile $\sim 6.4\kev{}$ and a possible Compton hump $\sim 10-20$\kev{}. These features clearly indicate the strong disc reflection and the presence of ionized absorbing material in the system. \citet{2006ApJ...652..603B} found the evidence of a broad iron emission feature at $\sim 6\kev{}$ and an iron absorption edge at $\sim 8\kev{}$ from the \rxte{} observation. In addition, an ionized Fe absorption line $\sim 6.96\kev{}$ has been reported by \citet{2012ApJ...753....2G}.

\begin{figure}
\includegraphics[scale=0.34, angle=-90]{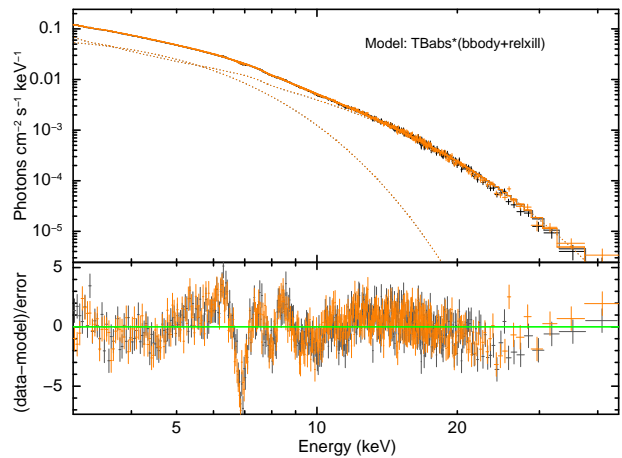}
\caption{ The unfolded spectral data, the best-fit model {\tt const*TBabs*(bbody+RELXILL)}. The lower panel of this plot shows the ratio of the data to the model in units of $\sigma$. Strong absorption features at $\sim 7$\kev{} and $\sim 8$\kev{} are clearly detected. } 
\label{Fig5}
\end{figure}

\subsection{Self-consistent reflection fitting}
When hard X-rays (Comptonized emission) irradiate the accretion disc, they produce a reflection spectrum that includes fluorescence lines, recombination, and other emission \citep{1989MNRAS.238..729F}. In most X-ray sources, the incident emission for the reflection spectrum is generally a hard power-law spectrum. However, in NS systems, the emission from the NS surface/boundary layer may be significant and contribute to the reflection \citep{2008ApJ...674..415C, 2019ApJ...873...99L, 2017ApJ...836..140L}. To reproduce both Comptonization and reflection spectra correctly, we replaced the {\tt cutoffpl} with {\tt RELXILL} \citep{2014ApJ...782...76G}. The model uses a cutoff power-law as an illuminating continuum for the reflection spectrum. This means that the component assumed to illuminate the accretion disc is consistent with the emergent reflection spectrum. The model parameters of {\tt RELXILL} are as follows: $q_{1}$ and $q_{2}$ are the inner and outer disc emissivity indices, respectively, $R_{break}$ is the break radius of the two emissivity indices, $R_{in}$ and $R_{out}$ are the inner and outer radii of the disc, respectively, $i$ is the inclination of the system, $a$ is the dimensionless spin parameter, $\Gamma$ is the photon index of the input cutoff power-law, log$\xi$ is the log of the ionization parameter, $A_{Fe}$ is the iron abundance of the system, $E_{cut}$ is the cutoff energy, $r_{refl}$ is the reflection fraction, and norm represents the normalization of the model. During the fitting with {\tt RELXILL}, we imposed the following: a single emissivity index $q_{1}=q_{2}=3$ (a value commonly found in X-ray binaries), a redshift of $z=0$ as it is a Galactic source, a spin of $a=0.25$ as the source has a spin frequency of $530$ Hz (considering $a\simeq 0.47/P_{ms}$ \citep{2000ApJ...531..447B} where $P_{ms}$ is the spin period in ms), and a large outer disc radius of $R_{out}=1000\;R_{g}$ (where $R_{g}=GM/c^2)$. The parameters $R_{in}$, $i$, $A_{Fe}$, log$\xi$, $\Gamma$, $E_{cut}$ were left free to vary. The addition of the model {\tt RELXILL} improved the fit significantly to $\chi^2/dof=1878/844$ ($\Delta\chi^2=-1834$ for the $5$ additional parameters). The corresponding spectrum for the model {\tt constant*TBabs*(bbody+RELXILL)} and the residuals are shown in Figure~\ref{Fig5}. The residuals showed a strong deviation from the model at $\sim 7$\kev{} and $\sim 8$\kev{}. These features suggest the presence of strong absorption lines  at $\sim 7$\kev{} and $\sim 8$\kev{} in the spectrum, as indicated by \citet{2006ApJ...652..603B}\\

\begin{figure*}
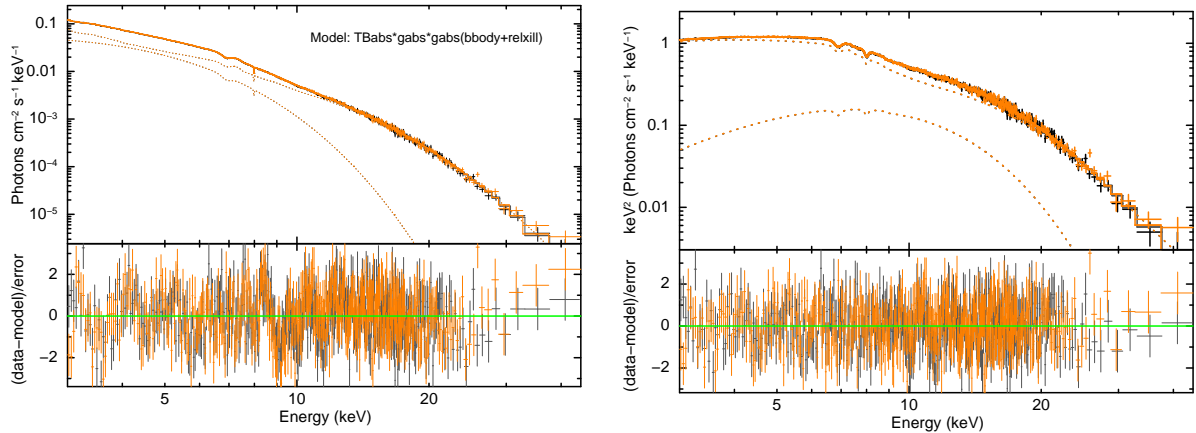

\includegraphics[scale=0.32, angle=-90]{fig6.ps}
\includegraphics[scale=0.32, angle=-90]{fig7.ps}
\caption{Left panel: The source spectrum is shown when the absorption features are fitted with two gaussian absorption models {\tt gabs} i.e.,{\tt const*TBabs*gabs*gabs(bbody+RELXILL)}. Right panel: Source spectrum after introducing one {\tt edge} component along with two {\tt gabs}, i.e.,{\tt const*TBabs*edge*gabs*gabs(bbody+RELXILL)}. The lower panels of each plot show the ratio of the data to the model in units of $\sigma$.} 
\label{Fig6}
\end{figure*}

\subsection{Residuals as absorption lines}
We identified strong and broad absorption lines at $\sim 7$\kev{} and a moderately broad absorption feature at $\sim 8$\kev{} in the spectrum. We tried to fit them with Gaussian absorption profiles {\tt gabs} in {\tt XSPEC}. Initially, we added two {\tt gabs} components, with all parameters free, to the existing model. The model {\tt constant*TBabs*gabs*gabs(bbody+RELXILL)} improved the fit markedly to $\chi^{2}/dof=1073/838$ (with $\Delta\chi^2=805$ for the $6$ additional parameters). The residual corresponding to this fit is shown in the left panel of Figure~\ref{Fig6}. The one Gaussian absorption feature has centroid energy centered at $E_{gabs}=6.90\pm 0.02$\kev{} and width $\sigma=0.13\pm 0.03$ \kev{}. The other Gaussian absorption line has centroid energy at $E_{gabs}=7.99\pm 0.03$\kev{} and width $\sigma=0.08_{-0.05}^{+0.07}$ \kev{}. However, some residuals still existed $\sim 9$\kev{} as shown in Figure~\ref{Fig6}. To model this feature  we added an absorption edge with energy and optical depth as free parameters. The result with the model {\tt constant*TBabs*gabs*gabs*edge(bbody+RELXILL)} was a further significant improvement in the fit to $\chi^{2}/dof=884/836$ (with $\Delta\chi^2=-189$ for the $2$ additional parameters). We measured the absorption edge energy at $8.79\pm 0.14$\kev{} and optical depth $0.038^{+0.022}_{-0.012}$. The best-fitting spectral parameters are listed in Table~\ref{parameters1}, and the associated residual is shown in the right panel of Figure~\ref{Fig6}.

\begin{figure}
\includegraphics[scale=0.34, angle=-90]{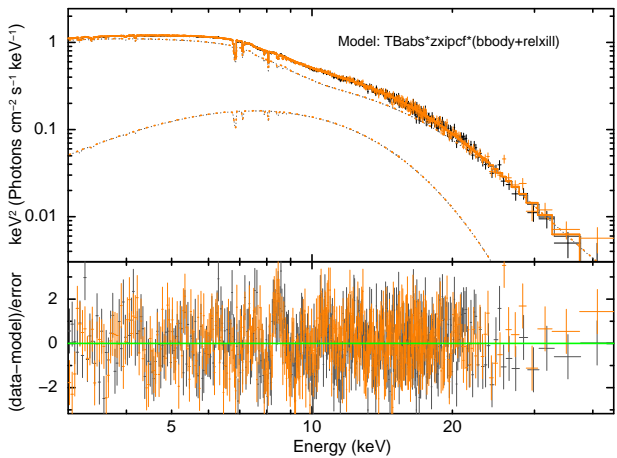}
\caption{The spectrum is fitted with the model {\tt const*TBabs*zxipcf(bbody+RELXILL)}. The lower panel of this plot shows the ratio of the data to the model in units of $\sigma$. The individual model components are also shown.} 
\label{Fig7}
\end{figure}

\subsection{Photo-ionized absorption}
\begin{figure*}
\includegraphics[scale=0.40, angle=0]{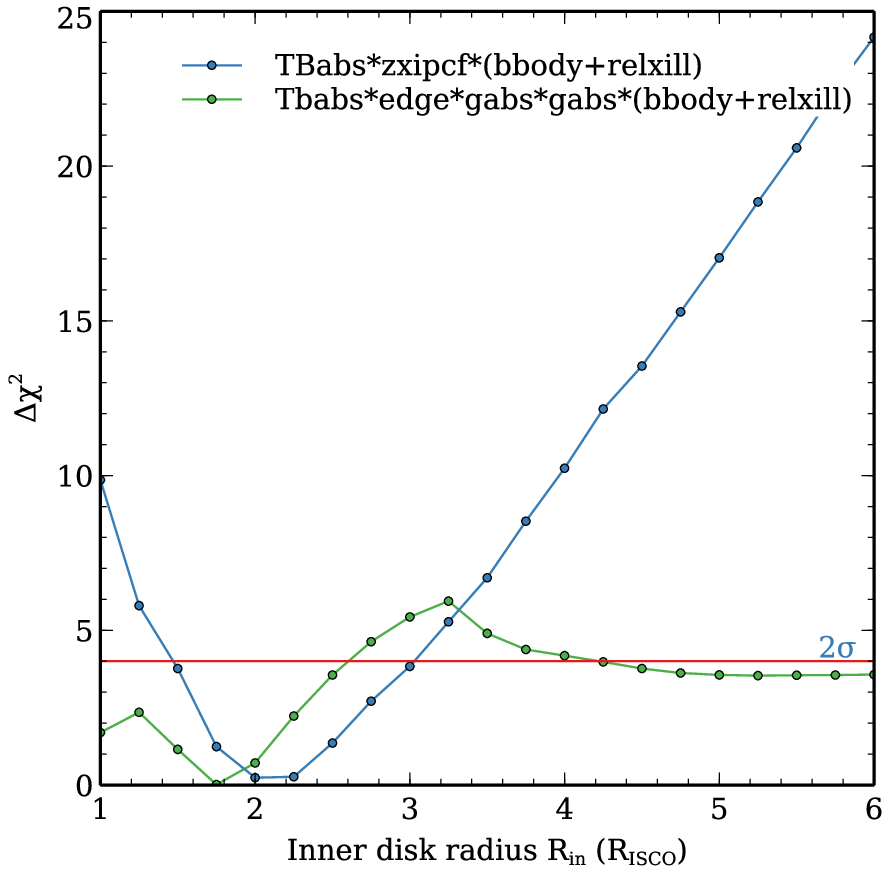}\hspace{2cm}
\includegraphics[scale=0.40, angle=0]{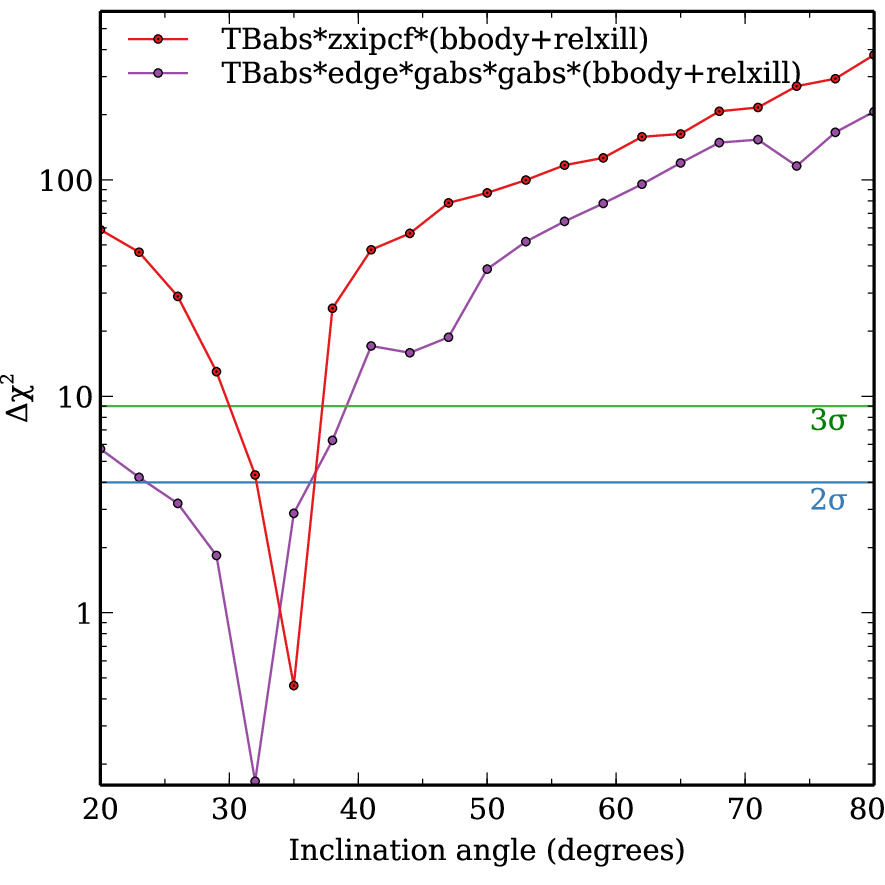}
\caption{The plots show the change in the goodness of fit for the inner disc radius ($R_{in}$) and disc inclination angle ($i$). The left panel shows the variation of $\Delta\chi^{2}(=\chi^{2}-\chi_{min}^{2})$ as a function of $R_{in}$ (varied between $1$ to $6\:R_{ISCO}$) obtained from two different model combination. The right panel shows the variation of $\Delta\chi^{2}(=\chi^{2}-\chi_{min}^{2})$ as a function of $i$ obtained from different models. We varied the disc inclination angle between $0$ degrees and $80$ degrees.} 
\label{Fig9}
\end{figure*}

Detection of several absorption features suggests the possibility that a local partially ionized absorber is located between the observer and the system as suggested by \citet{2005A&A...436..195B}. To verify this scenario, we replaced the Gaussian absorption profiles and absorption edge with a physically motivated warm absorber component {\tt zxipcf}, considering that the absorption lines are due to a photo-ionized plasma. The {\tt zxipcf} component uses a grid of XSTAR photoionized absorption models to describe the absorption of the incoming radiation by the plasma. The model considers a partial covering absorption characterized by covering fraction $f_{cov}$ of incoming photons by an ionized absorber whose ionization is described by log$\xi$. Besides these two parameters, the other main parameters of this model are the equivalent hydrogen column, $N_{H}$ and the redshift, $z$, of the absorbing material. We initially tried to fit this model, keeping all the parameters free except the redshift parameter $z$ (assumed zero shift). The model {\tt constant*TBabs*ZXIPCF*(bbody+RELXILL)} showed a drastic improvement in the fit, leading to $\chi^2/dof=1090/841$ (with $\Delta\chi^2=-788$ for the $3$ additional parameters). However, some residuals still existed in the $6.0-8.0$ \kev{} energy band, and we performed the fit again, considering $z$ as a free parameter. It significantly improved the fit to $\chi^2/dof=1002/840$ (with $\Delta\chi^2=-88$ for the $1$ additional parameter). We note that a model wherein the flow is required to have a redshift is significantly worse. Data and residual for this model are shown in Figure~\ref{Fig7}, and the best-fitting parameters of the {\tt zxipcf} model are reported in Table~\ref{parameters1}. The absorption features are well described with this warm absorber model. The rest of the parameter values obtained from this fit are consistent with those obtained earlier from the phenomenological fit with {\tt gabs} and absorption {\tt edge}. We note that similar fit statistics ($\chi^{2}_{\nu}\sim 1.28$) has been reported by \citet{2022ATel15429....1P} using similar combination of models for this source. We further used command {\tt steppar} in {\tt XSPEC} to search the best fit for $R_{in}$ and $i$ for the model {\tt constant*TBabs*gabs*gabs*edge(bbody+RELXILL)} and {\tt constant*TBabs*ZXIPCF*(bbody+RELXILL)}. The left and right panels in Figure~\ref{Fig9} show the $\Delta\chi^2$ of the fit versus the inner disc radius and the disc inclination for both models. \\

 \begin{table*}
   \centering
\caption{Fit results: Best-fitting spectral parameters of the \nustar{} observation of the source 1A~1744-361 using Model 1:  {\tt const*TBabs*gabs*gabs*edge(bbody+RELXILL)} and Model 2:{\tt const*TBabs*ZXIPCF(bbody+RELXILL)}} 
\begin{tabular}{|p{1.8cm}|p{4.8cm}|p{2.8cm}|p{2.8cm}}
    \hline
    Component     & Parameter (unit) & Model 1 & Model 2 \\
    \hline
    {\scshape Constant} & FPMB (wrt FPMA) & $1.006\pm 0.001$ & $1.006\pm 0.001$ \\
    {\scshape tbabs}    & $N_{H}$($\times 10^{22}\;\text{cm}^{-2}$) & $0.48_{-0.18}^{+0.13} $  & $0.34_{-0.12}^{+0.13} $   \\
    {\scshape zxipcf} & $N_{H,abs}$($\times 10^{22}\;\text{cm}^{-2}$)  &  --  & $2.94_{-0.42}^{+0.34}$ \\
    & log $\xi_{abs}$  & --  & $3.54_{-0.03}^{+0.08}$   \\
    & $f_{cov}$  & --  & $\lesssim 0.91$   \\
    & redshift ($z$) & -- & $-0.021_{-0.003}^{+0.001}$ \\
    {\scshape bbody} & $kT_{bb} (\kev)$ & $1.88_{-0.10}^{+0.06}$    & $1.87_{-0.05}^{+0.03}$ \\
    & Norm ($\times 10^{-3}$)& $4.00_{-0.72}^{+0.88}$   &  $4.15_{-0.59}^{+0.38}$ \\  
    {\scshape relxill} & $i$ (degrees) & $32_{-3}^{+5}$ & $35\pm 2$ \\
    & $R_{in}$($\times R_{ISCO}$) & $\lesssim 2.33$ & $2.00_{-0.39}^{+0.86}$\\
    & $\rm{log}\:\xi$(erg cm s$^{-1}$) &  $4.61_{-0.13}^{+0.05}$ &  $4.42_{-0.05}^{+0.09}$\\
    & $\Gamma$  & $1.17_{-0.04}^{+0.07} $ & $1.18_{-0.02}^{+0.05} $ \\
    & $A_{Fe}$ ($\times \;\text{solar})$   & $\lesssim 7.31$  & $6.40\pm 0.92$\\
    & $E_{cut}(\kev)$ &  $5.10_{-0.04}^{+0.07}$ & $5.11\pm 0.03$\\
    & $f_{refl}$   & $5.44_{-2.53}^{\dagger}$ & $6.31_{-0.93}^{+1.41}$ \\
    & norm ($\times 10^{-4}$)   &  $6.91\pm 0.01$ & $5.57_{-0.52}^{+2.80}$\\
    {\scshape gabs} &  $E_{gabs} (\kev)$ & $6.90\pm 0.02$ & -- \\
    & $\sigma_{gabs} (\kev)$ & $0.13\pm 0.03$ & -- \\
    & Strength ($\times 10^{-2}$) & $5.17\pm 0.01$ & -- \\
    {\scshape gabs} &  $E_{gbs} (\kev)$ & $7.99\pm 0.03$ & -- \\
    & $\sigma_{gabs} (\kev)$ & $0.08_{-0.05}^{+0.07}$ & -- \\
    & Strength ($\times 10^{-2}$) & $3.10_{-0.51}^{+1.10}$ & -- \\
    {\scshape edge} & $E_{edge}$ & $8.79_{-0.15}^{+0.12}$ & --\\
    & Optical depth $(\tau)$ & $0.04_{-0.01}^{+0.02}$ & -- \\
    {\scshape cflux} & $F_{bbody}^{*}$ ($\times 10^{-9}$ ergs/s/cm$^2$) & $0.34\pm 0.02$ & $0.36\pm 0.02$\\
    & $F_{relxill}^{*}$ ($\times 10^{-9}$ ergs/s/cm$^2$) & $2.01\pm 0.01$ & $1.99\pm 0.01$\\
    & $F_{total}^{*}$ ($\times 10^{-9}$ ergs/s/cm$^2$) & $2.35\pm 0.01$ & $2.35\pm 0.01$\\
   \hline 
    & $\chi^{2}/dof$ & $884/836$  & $1002/840$ \\
    \hline
  \end{tabular}\label{parameters1} \\
{\bf Note:} The outer radius of the {\tt RELXILL} spectral component was fixed to $1000\;R_{g}$. We fixed emissivity index $q1=q2=3$. The spin parameter ($a$) was fixed to $0.25$ as the spin frequency of the NS is $530$ Hz. $^{*}$All the unabsorbed fluxes are calculated in the energy band $3-70 \kev{}$ using the {\tt cflux} model component. $^{\dagger}$ upper bound error calculation is invalid.\\

\end{table*}

\section{Discussion}
We present the spectral and timing analysis results of the accreting neutron star 1A~1744-361 from the \nustar{} observation during its 2022 outbursts. During this observation, the source is detected with an average count rate of $\sim 80\ctssec$. A small flaring activity occured at $\sim 20$ ks after the start of the observation. The CCD indicates that the source was in the banana branch of the atoll track during this observation. The HR of the source lies in the range $0.05 - 0.15$, also indicative of the banana state of the source. The unabsorbed bolometric X-ray flux during this observation in the energy band $0.1-100\kev{}$ is $4.0\times 10^{-9}$ erg~s$^{-1}$ cm$^{-2}$. This implies an unabsorbed bolometric luminosity of $3.9\times 10^{37}$ erg~s$^{-1}$, assuming a distance of $9$ kpc. This value corresponds to $\sim 10\%$ of the Eddington luminosity ($L_{Edd}$) which is $\sim 3.8\times 10^{38}$ erg~s$^{-1}$ for a canonical $1.4\:M_{\odot}$ NS \citep{2003A&A...399..663K}. The luminosity is consistent with those of the atoll sources in the banana branch. All these observed characteristics indicate that the source was neither in the extremely low-hard state nor the very high-intensity (high-soft) state. We found that the continuum X-ray emission is well described by a blackbody ($kT_{bb}\sim 1.3\kev{}$) plus power-law ($\Gamma\sim 1.2$) with a cut-off energy at $\sim 5.0\kev{}$. The observed cut-off energy is typical for the atoll sources in the banana states. The spectrum of this source exhibited the presence of broad Fe-K emission line in $6-7\kev{}$, a Compton hump $10-20\kev{}$, and robust absorption features in $7-9\kev{}$. The broad Fe-K emission line and Compton hump are commonly explained by the disc reflection of the coronal emission (i.e., hard X-ray photons) subject to the strong relativistic distortion \citep{1989MNRAS.238..729F}. The strong absorption features indicate the presence of a local partially ionized absorber in the line of sight. The absorption by the plasma from the accretion disc generates absorption lines \citep{2009Natur.458..481N}. Therefore, in our spectral modeling, we employed disc reflection of hard X-ray emission and absorption of the incoming radiation in the photo-ionized plasma. \\

We found that the $3.0-50.0\kev{}$ source spectrum is adequately fitted using a model combination consisting of an absorbed single-temperature blackbody model ({\tt bbody}) and a reflection model ({\tt relxill}) along with the addition of a warm absorber model ({\tt zxipcf}). In this combination of model, the inner-disk radius, $R_{in}$, obtained from the {\tt relxill} is $\sim(1.61-2.86)R_{ISCO}=(8.4-14.9)R_{g}$ ($17.6-31.2$ km for a $1.4\Msun$ NS), where $R_{ISCO}=5.2R_{g}$ for a spinning NS with $a=0.25$). The source shows comparatively a larger inner disc radius and is consistent with other NS LMXBs (\citealt{2011ApJ...731L...7M, 2013MNRAS.429.3411P, 2016ApJ...819L..29K, 2016A&A...596A..21I, 2021MNRAS.504.1331M}). The value of $R_{in}$ indicates that the accretion disc is neither truncated very close to nor far from the NS  consistent with the observed flux/mass accretion rate. The photon index ($\Gamma=1.18\pm 0.03$) obtained from the final spectral fit further shows that the source is neither in the soft nor in the very hard spectral state. The best-fitting model gives a relatively low inclination $\sim 35^{0}$, not consistent with the dipping features previously observed in the light curve of this source. The reflection spectrum revealed that the accretion disc is highly ionized with log$\xi\sim 4.4$, and the iron abundance is high comparable to the solar composition ($A_{Fe}=6.40\pm 0.92$). The relativistic reflection model {\tt relxill} yielded values fully compatiable with those from the previous phenomenologica modeling with {\tt gabs} and {\tt edge}.\\

Apart from the disc reflection features, this observation showed the presence of prominent Fe absorption features between the energy ranges $6.9 - 8.8 \kev{}$ (see Figure~\ref{Fig5}). We used the Gaussian absorption model {\tt gabs} to estimate the features of the lines. The centroid energies of these lines are measured at $\sim 6.90\kev{}$ and $\sim 7.99\kev{}$. We further measured an absorption edge at energy $\sim 8.79\kev{}$. The strength of the absorption lines is measured in terms of the equivalent width (EW). The EW of the $\sim 6.90 \kev{}$ Fe line was $60\pm 2$ eV, and for the $\sim 7.99 \kev{}$ Fe line was $40\pm 1$ eV. The most abundant elements seen in absorption in the X-ray spectra of LMXBs are Fe XXV (rest-frame energy $6.70 \kev{}$) and Fe XXVI (rest-frame energy $6.97 \kev{}$), which can be red/blue shifted in case of an in/outflow. Thus, the line detected $\sim 6.90 \kev{}$ could represent absorption in a modest inﬂow or in a strong outﬂow. The in/outflow velocity ($v$) of the absorbing material can be determined via $(E_{0}-E_{i})/E_{i}=v/c$, where $E_{i}$ and $E_{0}$ are the observed and rest energy of the line, respectively. If the strong absorption feature we detect $\sim 6.90\kev{}$ is indeed the blueshifted Fe XXV K$\alpha$ line (rest energy $E_{0}=6.70 \kev{}$), then we measure an outflow velocity of $\sim 8700$ km s$^{-1}$. In addition, we estimated the full-width half-maximum (FWHM) of the intrinsic $\sim 6.90\kev{}$ absorption line of $\sim 0.306$ \kev{}. It corresponds to an equivalent resolved velocity of $\sim 13280$ km s$^{-1}$. However, considering $\sim 6.90$ \kev{} line as blueshifted Fe XXV predicts a strong Fe XXVI at commensurate blueshift ($\sim 7.18$ \kev{}), which is not seen. Thus, the observed line at $\sim 7.99 \kev{}$ probably is the blueshifted line from the K$\beta$ transition of He-like, Fe XXV ion, compatible with rest energy $\sim 7.78 \kev{}$. We found that the absorption edge at energy $\sim 8.79\kev{}$ is consistent with a He-like Fe XXV absorption edge.\\

The prominent absorption features are suggestive of the presence of a local partially ionized absorber. For this reason, we implemented the model {\tt zxipcf} that accounts for the contribution of the partially ionized local absorber. Moreover, the line-by-line fitting is not sufficient to address the question of inflow/outflow to the system; a physical self-consistency is required \citep{2016ApJ...822L..18M}. We found that the partially covering absorber has a significant absorbing column density of $N_{H}=2.94_{-0.42}^{+0.34}\times 10^{22}$ cm$^{-2}$. The absorbing material covers a fraction of $\sim 90\%$ of the central X-ray source, characterized by a high ionization parameter (log$\xi\sim 3.6$). The value of the ionization parameter is consistent with \citet{2012ApJ...753....2G}, determined by the XSTAR simulations. This model further gives a redshift of $z=-0.021_{-0.003}^{+0.001}$, i.e. implying an outflow velocity of $\simeq 6000-7200$ km s$^{-1}$. This indicates that the absorption is best associated with a moderate outflow from the NS system.\\ 

We note that relativistic X-ray reflection and photoionized absorption have been simultaneously detected in bright {\it Z}-type NS LMXB GX~13+1 (\citealt{2004ApJ...609..325U, 2012A&A...543A..50D, 2023MNRAS.522.3367S}). The spectral behavior of the source 1A~1744-361 is very similar, as observed for the atoll source GX~13+1 by \citet{2023MNRAS.522.3367S}. They detected broad ($\sigma\sim 0.15$) absorption lines around $\sim 6.87$ and $\sim 8.03\kev{}$ from the \nustar{} spectrum of GX~13+1 and attributed those to the Fe XXVI K$\alpha$ and Ni XXVIII absorption line, respectively. If the observed line energies are compared to the rest energies of Fe XXVI K$\alpha$ and Ni XXVIII absorption line, it indicates a redshifted flow. However, from the modeling with a physically motivated warm absorber model, they suggested an outflow from the system with a velocity $\lesssim 1000$ km s$^{-1}$. They attributed the absorption features to a photo-ionized plasma emanating from the accretion disc in the form of wind/jets. Blueshifted narrow absorption lines have been previously observed in many NS LMXBs and are interpreted as outflowing disc winds \citep{2004ApJ...609..325U, 2011ApJ...731L...7M, 2014ApJ...796L...9D, 2016A&A...589A.102B}. The amount of blueshift ($\sim 6300$ km s$^{-1}$) observed for the source 1A~1744-361 is comparatively larger than the typically observed values i.e., $400-3000$ km s$^{-1}$ \citep{2016AN....337..368D}. However, the highest velocities claimed for NS LMXBs so far are $\sim 9000-14000$ km s$^{-1}$ \citep{2014ApJ...796L...9D, 2016ApJ...822L..18M}. For the NS LMXB 1RXS~J180408.9-342058, \citet{2016MNRAS.461.4049D} measured an outflow velocity as high as $\sim 25800$ km s$^{-1}$ even at a low inclination angle of $i\sim 30^{0}$. Whereas redshifted flow, i.e., inward to the NS systems, is not commonly observed, and the origin of the same is still poorly understood. \citet{2018ApJ...868L..26K} reported a redshift of $\sim 4000\pm 1400$ km s$^{-1}$ for the high-mass X-ray binary SMC X-1 and discussed some possibilities of the astrophysical origin of the absorption lines. \\

A rough estimation of the distance of the ionized absorbing material ($r$) from the centre of the system could be evaluated by means of the ionization parameter $\xi$. According to \citet{2012A&A...543A..50D}, the distance $r$ is defined as 
\begin{equation}
r=\frac{L}{\xi\:N_{H,warmabs}}\frac{d}{r},
\end{equation} 
where $L$ is the unabsorbed luminosity in ergs s$^{-1}$, $n_{e}=N_{H, warmabs}/d$ is the electron density of the ionized plasma, and $d$ is the thickness of the slab of ionizing absorbing material. Using the spectral-fitting results and considering typical values of $d/r$ ranging from $0.1$ to $1$, we estimate $r$ between $\sim 1.0\times (10^5-10^6)$ km. This value is in excellent agreement with the value previously obtained by \citet{2012ApJ...753....2G} using the \chandra{} data. Moreover, this value for $r$ is safely within the accretion disc radius found earlier by \citet{2012ApJ...753....2G}. It suggests that the absorption features are most likely emanating from a disc wind far from the central source. However, the complete geometry of the putative absorber in 1A~1744-361 needs to be clearly understood.\\

The best-fit spectral parameters were further used to compute physical properties like mass accretion rate ($\dot{m}$), the maximum radius of the boundary layer ($R_{BL,max}$), and magnetic field strength ($B$) of the NS in the system. We first estimated the mass accretion rate per unit area, using Equation (2) of \citet{2008ApJS..179..360G}
\begin{equation}
\begin{split}
\dot{m}=&\:6.7\times 10^{3}\left(\frac{F_{p}\:c_\text{bol}}{10^{-9} \text{erg}\: \text{cm}^{-2}\: \text{s}^{-1}}\right) \left(\frac{d}{10 \:\text{kpc}}\right)^{2} \left(\frac{M_\text{NS}}{1.4 M_{\odot}}\right)^{-1}\\
 &\times\left(\frac{1+z}{1.31}\right) \left(\frac{R_\text{NS}}{10\:\text{km}}\right)^{-1} \text{g}\: \text{cm}^{-2}\: \text{s}^{-1}.
 \end{split} 
\end{equation}
The above equation yields a mass accretion rate of $4.4\times 10^{-9}\;M_{\odot}\;\text{y}^{-1}$ at a persistent flux $F_{p}=2.35\times 10^{-9}$ erg~s$^{-1}$ cm$^{-2}$, assuming the bolometric correction $c_{bol} \sim 1.38$ for the nonpulsing sources \citep{2008ApJS..179..360G}. Here we assume $1+z=1.31$ (where $z$ is the surface redshift) for an NS with mass ($M_{NS}$) 1.4 $M_{\odot}$ and radius ($R_{NS}$) $10$ km. To estimate the maximum radial extension of the boundary layer from the NS surface based on the mass accretion rate, we used Equation (2) of \citet{2001ApJ...547..355P}. We found the maximum value of the boundary layer to extend to $R_{BL}\sim 6.53\;R_{g}$, assuming $M_{NS}=1.4\:M_{\odot}$ and $R_{NS}=10$ km. The actual value may be larger than this if we account for the viscous effects and the spin of this layer. The radial extent of the boundary layer regions is consistent with the disc position. \\

An upper limit of the magnetic field strength of the NS can be estimated using the upper limit of $R_{in}$ measured from the reflection fit. Equation (1) of \citet{2009ApJ...694L..21C} gives the following expression for the magnetic dipole moment,
\begin{equation}
\begin{split}
\mu=&3.5\times 10^{23}k_{A}^{-7/4} x^{7/4} \left(\frac{M}{1.4 M_{\odot}}\right)^{2}\\
 &\times\left(\frac{f_{ang}}{\eta}\frac{F_{bol}}{10^{-9} \text{erg}\: \text{cm}^{-2}\: \text{s}^{-1}}\right)^{1/2}
 \frac{D}{3.5\: \text{kpc}} \text{G}\; \text{cm}^{3}.
\end{split} 
\end{equation}
We assumed a geometrical coefficient $k_{A}=1$, an anisotropy correction factor $f_{ang}=1$, and the accretion efficiency in the Schwarzschild metric $\eta=0.2$ (as reported in \citealt{2009ApJ...694L..21C} and \citealt{2000AstL...26..699S}) and used $0.1 - 100$\kev{} flux as the bolometric flux ($F_{bol}$) of $\sim 4.0\times 10^{-9}$ erg~s$^{-1}$ cm$^{-2}$. \citet{2009ApJ...694L..21C} modified $R_{in}$ as $R_{in}=x\:GM/c^{2}$ introducing a scale factor $x$. Using the upper limit of $R_{in}$ ($\lesssim 15\:R_{g}$), we obtained $\mu \leq 0.47\times 10^{27}$ G cm$^{3}$. This corresponds to an upper limit of the magnetic field strength of $B\lesssim 0.94\times 10^{9}$ G at the magnetic poles (for an NS mass of $1.4\:M_{\odot}$, a radius of $10$ km, and a distance of $9$ kpc). To the best of our knowledge, there has not been any previous estimation of the $R_{in}$, $\dot{m}$, $\mu$, $B$ of this NS system. Hence, there exists no scope for comparison of the results obtained in the present analysis with other previous works. \\

In this work, we aimed to focus on detecting disc reflection and absorption, thereby constraining its accretion geometry through spectro-timing analysis. We note that the continuum emission can alternatively be described with  different combinations of model components. Apart from the presence of the Comptonization component, previous studies detected the presence of a broad emission line and absorption features in the spectra of 1A~1744-361 \citep{2006ApJ...652..603B, 2012ApJ...753....2G} and we focused on these features. We primarily inspected that the reflection and absorption features were insensitive to the specific choice of the continuum models. However, the choice of reflection and absorption model may have some impact on the spectral parameters. In our spectral fitting, we have found that power law is dominant in the phenomenological fit and provided most of the hard X-ray flux. This flux illuminates the accretion disc and produces the reflection spectrum. The choice of the reflection model is based on this observed fact. However, different reflection models based on different illuminating spectra may be tested in the reflection fits. We note that it is challenging to constrain all the reflection parameters if we implement the reflection models, which assumes different illuminating spectra. In addition, a further correlation between the ionization and column density of the absorber and the equivalent width of the broad iron line can be drawn in future studies.

\section{Acknowledgements}
This research has made use of data and/or software provided by the High Energy Astrophysics Science Archive Research Centre (HEASARC). This research also has made use of the \nustar{} data analysis software ({\tt NuSTARDAS}) jointly developed by the ASI Space Science Data Center (SSDC, Italy) and the California Institute of Technology (Caltech, USA). ASM and BR would like to thank Inter-University Centre for Astronomy and Astrophysics (IUCAA) for their facilities extended to him under their Visiting Associate Programme.

\section{Data availability}
This research has made use of data obtained from the HEASARC, provided by NASA's Goddard Space Flight Center. The observational data set with Obs. IDs $90801312001$ (\nustar{}) dated 8th June 2022 is in public domain put by NASA at their website https://heasarc.gsfc.nasa.gov.

\def\apj{ApJ}
\def\apjl{ApJl}
\def\pasp{PASP} \def\mnras{MNRAS} \def\aap{A\&A} \def\physerp{PhR} \def\apjs{ApJS} \def\pasa{PASA}
\def\pasj{PASJ} \def\nat{Nature} \def\memsai{MmSAI} \def\araa{ARAA} \def\iaucirc{IAUC} \def\aj{AJ} \def\aaps{A\&AS}
\bibliographystyle{mn2e}
\bibliography{aditya}

\end{document}